\documentstyle[12pt]{article}
\newcommand{\ba}{\begin{eqnarray}}
\newcommand{\ea}{\end{eqnarray}}
\newcommand{\be}{\begin{equation}}
\newcommand{\ee}{\end{equation}}
\newcommand{\ed}{\end{document}}
\newcommand{\lab}[1]{\label{#1}}
\newcommand{\re}[1]{(\ref{#1})}
\newcommand{\ci}[1]{\cite{#1}}
\renewcommand{\baselinestretch}{1.2}
\date{}
\title{ WAVE FUNCTIONS AND ENERGY TERMS OF THE SCHR\"ODINGER EQUATION \\
WITH TWO-CENTER COULOMB PLUS  HARMONIC OSCILLATOR POTENTIAL}
\author{ D.U.MATRASULOV \\
 Heat Physics Department of the Uzbek Academy of Sciences,\\
28 Katartal St.,700135 Tashkent, Uzbekistan }
\begin{document}\large
\maketitle
\begin{abstract}
Schr\"odinger equation for two center Coulomb plus harmonic oscillator
potential is solved by the method of ethalon equation at
large intercenter separations. Asymptotical expansions for energy term and
wave function are obtained in the analytical form.

\end{abstract}
PACS numbers: 03.65.Ge, 03.65 -W, 12.39.Pn\\

The Schr\"odinger equation with two-center potentials is of considerable
interest in various problems related with few body systems treating in
Born-Oppenheimer approximation. It describes the bound states of light
particle in the field of two heavy particles. Usually such type of systems
arise in molecular physics. However last years there is a considerable
interest to another systems modelling by two-center Schr\"odinger equations,
namely baryons containing two heavy quarks (QQq- baryons)\ci{Rich92} and heavy flavoured
hybryd mesons (QQg-mesons) \ci{Mukh93} are becoming subject of extensive investigation now.
The motion of light quark (gluon) in the field of two heavy quarks
(quark-antiquark pair) can be described in the nonrelativistic approach by the
Schr\"odinger equation with two-center Coulomb plus confinement potential.
In this paper we concerned with the Schr\"odinger equation with two-center
Coulomb plus harmonic oscillator potential which is, to our knowledge, the only
Coulomb plus confinement potential allowing
separation of variables in the Schr\"odinger equation. Using the method
of ethalon equation, widely applying for the solution of two-center
Coulomb Schr\"odinger equation in molecular physics, we obtain wave
functions and eigenvalues of the Schr\"odinger equation with two-center Coulomb
plus harmonic oscillator potential in the form of asymptotical expansion in terms
of inverse powers of the intercenter distance.

So we want to find energy eigenvalues and wave functions of the following
Schr\"odinger equation
\be
[-\frac{1}{2}\Delta -
\frac{Z}{r_{1}}-\frac{Z}{r_{2}}+\omega^{2}(r_{1}^2+r_{2}^2)]\psi=E\psi \ee

In the prolate spheroidal coordinates, which are defined as follows:
$$\xi=\frac{r_{1}+r_{2}}{R}\,\, (1<\xi<\infty),\,\,\,
 \eta=\frac{r_{1}-r_{2}}{R}\,\, (-1<\eta<1)$$
this potential can be written in the form
\be V(r_{1},r_{2})= -\frac{2}{R^2}
\frac{a(\xi)+b(\eta)}{\xi^2-\eta^2}+\frac{\omega^2R^2}{2}
\lab{pot}
\ee
where
$$a(\xi)=2ZR-\frac{\omega^2R^4}{4}\xi^2(\xi^2-1),\,\,
b(\eta)=2ZR-\frac{\omega^2R^4}{4}\eta^2(\eta^2-1)$$

As is well known \ci{Kom78}, the Schr\"odinger equation with potential in
the form \re{pot} is separable in the prolate spheroidal coordinates.
After separation we have three ordinary differential equations connected
by separation constants $\lambda$ and $m$:
$$
[\frac{\partial}{\partial\xi}(\xi^2-1)\frac{\partial}{\partial\xi}+a\xi+
(p^2-\gamma'\xi^2)(\xi^2-1)-\frac{m^2}{(\xi^2-1)}-\lambda]X(\xi)=0  $$

$$
[\frac{\partial}{\partial\eta}(\eta^2-1)\frac{\partial}{\partial\eta}+
(p^2-\gamma'\eta^2)(\eta^2-1)-\frac{m^2}{(\eta^2-1)}-\lambda]Y(\eta)=0
$$ $$
(\frac{\partial^2}{\partial\phi^2}+m^2)Z(\phi)$$
where
$$p=\frac{R}{2}\sqrt{2E'}$$
$$ E' = E -\frac{\omega^2R^2}{2}\;\;\;
\gamma'=\frac{\omega^2R^4}{4},\;\;\;a=2ZR.
$$

Boundary conditions for $U$ and $V$ are
\be U(\xi)\mid_{\xi=1} =0 ,\;\;\; U(\xi)\mid_{\xi \longrightarrow \infty}
\longrightarrow 0 ,
\lab{radbound}
\ee
\be V(\eta)\mid_{\eta =\pm 1} = 0.
\lab{angbound}
\ee

 After substitutions $$ U(\xi)=\frac{1}{\sqrt{\xi^2-1}}X(\xi)$$ $$
V(\eta)=\frac{1}{\sqrt{1-\eta^2}}Y(\eta),$$ these equs. can be reduced
to the following canonical form:

\be
U''(\xi)+[\frac{h^2}{4}+\frac{h(\alpha\xi-\lambda)}{\xi^2-1}-h^4\gamma\xi^2
+\frac{1-m^2}{(\xi^2-1)^2}]U(\xi)=0
\lab{rad}
\ee
\be
V''(\eta)+[\frac{h^2}{4}+\frac{h\lambda}{1-\eta^2}-h^4\gamma\eta^2
+\frac{1-m^2}{(1-\eta^2)^2}]V(\eta)=0
\lab{ang}
\ee
where $\alpha=2Z/\sqrt{2E'},$ $\gamma=\omega^2/8E'^2 $ ,
\be
h=2p
\lab{energy}
\ee

\section{Asymptotics of quasi-angular equation.}
We will solve eqs.\re{rad} and\re{ang} for large $R$ approximately by the
method of ethalon equation. This method is sucesfully applied to the
solution of nonrelativistic two center Coulomb problem \ci{Kom78,Slav,Kom67} and in
the theory of difraction of waves.  Details of the method of
ethalon equation are given in \ci{Kom78,Slav,Kom68,Kom67}.  Let's start from the
angular eq.\re{ang}. As an ethalon equation for eq.\re{ang} we choose the
Whittaker equation \ci{Erd}:  \be
 W''+[-\frac{h^4}{4}+\frac{h^2k}{z}+\frac{1-m^2}{4z^2}]W \lab{ethalon1} \ee
and seek solution  in the form
\be V=[z'(\eta)]^{-\frac{1}{2}}M_{k,\frac{m}{2}}(h^2z),
\lab{sol1}
\ee
where $M_{k,\frac{m}{2}}(h^2z)$ is the solution(regular at zero) of
eq.\re{ethalon1}. After substitution \re{sol1} into \re{ang} we get
following equation for $z$:
\ba
\frac{z'^2}{4}-\gamma(x-1)^2 - \frac{1}{h^2}(\frac{1}{4}
+\frac{kz'^2}{z} -\frac{\lambda}{2x(1-x/2)})+
\nonumber \\
\frac{\tau}{h^2}(\frac{1}{x^2(1-x^2)} -\frac{z'^2}{z^2})
-\frac{1}{2h^2}[z,x]=0
\lab{zeq}
\ea
where $\tau=\frac{1-m^2}{4},$ $x=1+\eta$. Requirement  coincidence
of trnsition points
$$ z(x)\mid_{x=0}=0 $$
leads to the following "quantum condition"

\be \lambda = 2kz'(0)+ \frac{2\tau}{h^2}[\frac{z''(0)}{z'(0)} -1]
\lab{lambda1}
\ee
We will seek the solution of eq.\re{zeq} and eigenvalues $\lambda$ in the
form of following asymptotical expansion:  $$ z=\sum^{\infty}_{k=0}
\frac{z_{k}}{h^k},\;\;\;\;\;
\lambda=\sum^{\infty}_{k=0}\frac{\lambda_{k}}{h^k} $$

Substitution these expansions into \re{zeq} gives us the recurrence
system of differentil equations for $z$ :
$$z'_{0}=2\gamma^{\frac{1}{2}}(x-1) $$
$$z'_{1}=0 $$
$$ z'_{2}=\frac{1}{2z'_{0}}+\frac{2kz'_{0}}{z_{0}} -
\frac{(z'_{1})^2}{2z'_0} - \frac{2\lambda_0}{z'_0x(1-x/2)} -
\frac{z'^2_2}{2} $$
$$ .\; .\; .\; .\; .\; .\; .\; .\; .\; .\; .\; .\; .\; .\; .\; .\; .\; .$$
and for $\lambda$ :  $$\lambda_0 =2kz'_0(0)$$
$$\lambda_1 = 2kz'_1(0) $$ $$\lambda_2 = 2kz'_2(0) +
2\tau(\frac{z''_0(0)}{z'_1(0)} - 1)$$ $$ .\;\; .\; .\; .\; .\; .\; .\; .\;
.\; .\; .\; .\; .\; .\; .$$

Solving these recurrence equations we have for $\lambda$ :
\be \lambda^{(\eta)} = 4k\gamma^{\frac{1}{2}} + \frac{2k\beta - 4\tau}{h^2}
+ O(\frac{1}{h^4})
\lab{lambda1}
\ee

and for $z$ :
\be z= \gamma^{\frac{1}{2}}x(2-x) + \frac{1}{h^2}\beta ln(1-x)
+ O(\frac{1}{h^4})
\lab{wave1}
\ee

From boundary conditions one can obtain for quantum number $k$ \ci{Kom68,Kom67}:
$$ k = q + \frac{m+1}{2}, $$  where $q = 0,1,2,,... $.

\section{Asymptotics of quasi-radial equation}
For equation \re{rad} as an ethalon equation we take
equation :
\be W''+[h^2s - h^4y^2 - \frac{4\tau+3}{4y^2}]W =0 \ee
solution of which expressed by the confluent hypergeometric
functions\ci{6,7}
$$ W = y^c e^{-\frac{h^4y^2}{2}}
F(\frac{s-2c -1}{4}, c+\frac{1}{2}, h^4y^2)$$

where $c = \frac{1+\sqrt{m^2+3}}{2}$.

Boundary condition \re {radbound} and properties of functions $F$
\ci{Erd,Abr} give rise to the  following expression for s:
 $$ s = 4n + \sqrt{m^2 +3} +2 .$$

Substituting $$ U = [y(\xi)]^{-\frac{1}{2}}W(y(\xi)) $$ into
eq.\re{rad} we have
\ba
\frac{y^2y'^2}{4} - \gamma\xi^2 +
\frac{1}{ h^2}(\frac{1}{4} - sy'^2 - \frac{\lambda}{\xi^2 -1})
+\frac{1}{h^3} \frac{\alpha\xi}{\xi^2 -1} +
\nonumber \\
\frac{4\tau}{h^4(\xi^2-1)^2} -
\frac{3-4\tau}{4h^4}\frac{y'^2}{y^2} -\frac{1}{2h^4} [y,\xi] = 0.
\lab{nonr}
\ea
After substitution $$\phi = \frac{y^2(t)}{4}$$ this eq. can be reduced
to the form

\ba
\phi'^2 -\gamma(t+1) +\frac{1}{h^2}(\frac{1}{4}-(n+\frac{1}{2})
\frac{\phi'^2}{\phi}-\frac{\lambda}{t(t+2)})+\frac{1}{h^3}\;\frac{\alpha
(t+1)}{t(t+2)} \nonumber \\ +\frac{\tau}{h^4}(\frac{\phi'^2}{\phi^2} -
\frac{4}{t^2(t+2)^2}) - [\phi,t]= 0.
\lab{nonlin}
\ea
where $t = \xi -1$
Quantization condition which follows from $\phi(x) =0$ is written in the
form
\be
\lambda = -2s\phi'(0) + \frac{\alpha}{h} -\frac{1}{h^2}
[\frac{\phi''}{\phi'} + 1]\mid_{t=0}
\ee

Inserting into eq.\re{nonlin} asymptotical expansions
$$ \phi=\sum^{\infty}_{k=0} \frac{\phi_{k}}{h^k},\;\;\;\;\;
\lambda=\sum^{\infty}_{k=0}\frac{\lambda_{k}}{h^k} $$
and solving obtained equations we have for $y$:
\ba
y = 2\gamma^{\frac{1}{4}}(t^2+2t)^{\frac{1}{2}} + \frac{1}{h^2}
\delta\gamma^{-\frac{1}{4}}(t^2+2t)^{-\frac{1}{2}}ln(t+1) +
\nonumber \\
\frac{1}{h^3}\alpha\gamma^{-\frac{3}{4}}(t^2+2t)^{-\frac{1}{2}}ln\frac{2(t+1)}{t+1}
+ O(\frac{1}{h^4})
\lab{wave2}
\ea

and for $\lambda$
\be \lambda^{(\xi)} = -2s\gamma^{\frac{1}{2}} - \frac{\alpha}{h} +
\frac{4\tau -s\delta}{h^2} - \frac{s\alpha \gamma^{-\frac{1}{4}}}{2h^3} +
 O(\frac{1}{h^4})
\lab{lambda2}
\ee
\section{Asymptotical expansion for energy and wave functions.}
 Asymptotical expansions \re{lambda1} and \re{lambda2} give us expression
 for energy in the form of multipole expansion.
In order to obtain this expansion one should insert
$$ E' = E_{0} + \frac{E_{1}}{R} + \frac{E_{2}}{R^2} + ...  $$
into \re{lambda1} and \re{lambda2}.
Equating $\lambda^{\eta}$  and  $\lambda^{\xi}$ and taking into account
\re{energy} we get following equations for coeficients $E_{1}, E_{2},
 ... $ :

$$ E_{1} = \frac{1}{6Z}[(s\omega-2k\omega^{-1})(2E_{0})^{\frac{5}{2}} +
 (4s^2 -16k^2 -16\tau)(2E_{0})^{\frac{3}{2}}] , $$
$$ E_{2} = \frac{5}{2}E_{1}^2 + 2s\omega^{-1}E_{0} +
E_{1}(2E_{0})^{\frac{1}{2}}Z^{-1}(16\tau^2 + 16k^2 - 4s^2), $$
$$ .\,.\,.\,.\,.\,.\,.\,.\,.\,.\,.\,.\,.\,.$$

Now we need to find $E_{0}$. In order to find this one we note that
for $ R \longrightarrow \infty $  $E' = E_{0}$ or
\be E = E_{0} + \frac{\omega^2R^2}{2}.
\lab {energy2}
\ee
On the other hand for large $R$
\ba
V(r_{1}, r_{2}) = \frac{2Z}{R}\sum^{\infty}_{l=0}(\frac{r}{R})^l
 P_{l}(cos\theta) + \omega^2[r^2 +2rRcos\theta +\frac{R^2}{4}) +
\nonumber \\
(r^2 - 2rRcos\theta +\frac{R^2}{4})]\approx
 \omega^2(2r^2 + \frac{R^2}{2}).
\ea

Hence for the energy term with this potential  we have
\be E = 2\omega(N + \frac{3}{2}) + \frac{\omega^2R^2}{2},
\lab{energy3}
\ee
 where $N = 2n +q + m $ is the principial quantum number. Comparing
\re{energy2} and \re{energy3} we have
$$ E_{0} = 2\omega(N +\frac{3}{2}).$$

Thus we have obtained an asymptotical expansion for
wave functions and energy eigenvalues of the Schr\"odinger
equation with two-center Coulomb plus harmonic oscillator potential.
Derived formulas have to be useful for farther
numerical calculations in nonasymptotical region and can be also used for
estimation QQq baryon energy spectra.

\ed